\documentstyle[11pt,newpasp,twoside,epsf]{article}
\def\edcomment#1{\iffalse\marginpar{\raggedright\sl#1\/}\else\relax\fi}
\reversemarginpar

\begin{document}
\title{Effects of Photospheric Temperature Inhomogeneities on Lithium 
       Abundance Determinations (2D)}
\author{Roger Cayrel}
\affil{Observatoire de Paris, DASGAL, 61, Observatoire de Paris \\
       F75014 Paris, France}
\author{Matthias Steffen}
\affil{Astrophysikalisches Institut Potsdam,  An der Sternwarte 16 \\ 
       D-14482 Potsdam, Germany }

\begin{abstract} Based on detailed 2D radiation hydrodynamics (RHD) 
simulations, we have investigated the effects of photospheric temperature 
inhomogeneities induced by convection on spectroscopic determinations of
the lithium abundance.
Computations have been performed both for the solar case and for a metal-poor 
dwarf. NLTE effects are taken into account, using a five-level atomic model for
Li~I. Comparisons are presented with traditional 1D models having the same 
effective temperature and gravity. The net result is that, while LTE results 
differ dramatically between 1D and 2D models, especially in the metal-poor
case, this does not remain true when NLTE effects are included: 1D/2D
differences in the inferred NLTE Li abundance are always well below 0.1 dex. 
The present computations still assume LTE in the continuum. New computations 
removing this assumption are planned for the near future.  
\end{abstract}

\section{Introduction}

 It would be an offense to the audience here to pretend to explain 
 why it is important to determine accurately the abundance of $^6$Li and $^7$Li
 in the oldest stars. In this respect, we have nothing to add to the exposition
 by Fran\c{c}ois Spite (this volume). We will right away mention the two major 
 problems which may cast doubts on our real knowledge of the actual initial 
 abundance of Li in the oldest stars, and consequently in the primordial matter.
 (i) While standard models of the internal structure of metal-poor dwarfs do 
 not deplete $^7$Li, more sophisticated models including rotationally induced 
 mixing (Pinsonneault et al.\ 1992) have  predicted that the measured abundance 
 in the photosphere is 5 to 10 times less than the initial abundance 
 representative of Big Bang material. (ii) On top of that, Kurucz (1995) 
 claimed that the hot and cold convective structures produce large effects in 
 metal-poor stellar photospheres, where the convection zone reaches the 
 line formation layers. The claimed effect is an overionization of Li by a 
 factor of 10, leading to an underestimation of the abundance of Li when 
 derived from the resonance line of Li~I ($\lambda$~670.8~nm) in the usual way.
 
 If these two statements are correct, the true abundance of Li in primordial 
 matter is 50 to 100 times higher than the value derived from 1D, LTE models 
 of halo subdwarfs so far. The first factor of 5 to 10 has been discussed in a 
 previous paper by Ryan (this symposium), and shown to be likely much smaller, 
 of the order of 1 to 1.4. We shall not come back to this point, which we 
 consider as very well treated.  
 
 Before this symposium, a single paper (Asplund et al.\ 1999) has dealt with 
 the question of the other factor of 10 claimed by Kurucz (1995), whose 
 arguments were based on a simplified two-column model.
 In contrast, the work by Asplund et al.\ relies on realistic 3D hydrodynamical 
 models, similar to the simulations of the solar granulation (Stein \& Nordlund 
 1998), but with parameters appropriate for two metal-poor stars: HD 140283 and 
 HD 84937, both subgiants. The computation of the lithium resonance line was 
 made under the assumption of LTE, and the correction to be applied to the Li
 abundance derived from standard 1D models was found to be large, of the order 
 of -0.2 to -0.35 dex. Note that these corrections have the opposite sign as 
 Kurucz's prediction!
 However, Kiselman (1997, 1998) had shown, in the solar case, that NLTE and LTE 
 computations lead to significantly different values of equivalent widths of 
 the Li~I $\lambda$~670.8~nm line over hot and cold structures (see Fig.~3 
 of his 1997 paper, top panel). 
 
 For this reason, we decided to undertake NLTE 
 radiation hydrodynamics computations for the case of a metal-poor star, and we 
 report here on the results of this investigation. In the next section we recall
 former work related to simulations of the solar granulation, a useful benchmark
 for checking the theory, but not directly applicable to metal-poor stars. 
 In section~3 we describe the assumptions underlying the construction of the 
 2D RHD models used for the spatially resolved computation of the lithium 
 resonance line. Section~4 gives the description of the NLTE treatment
 of the Li atom, and section~5 summarizes our results and compares them to 
 those presented by M. Asplund (this symposium). Finally, our conclusions are 
 listed in section~6.

\section{Former work at solar metallicity}

While there is only one paper dealing with multidimensional atmospheres for 
metal-poor stars (cited above), there are several studies for the solar case, 
aimed at understanding the variation of the continuous radiation intensity 
(granulation), and the behavior of spectral lines across the solar granulation 
pattern. Several of these works use snapshots from 3D simulations by Stein \& 
Nordlund (1998), such as Kiselman (1997,1998) and Uitenbroek (1998). Gadun \& 
Pavlenko (1997) use their own 2D simulations.
 
Of particular interest for us are the papers dealing with the combined effects 
of multidimensional structures and NLTE (Kiselman and Uitenbroek). It is clear 
from Kiselman (1997) that the NLTE behavior of the Li~I $\lambda$~670.8~nm 
resonance line is drastically different from its LTE behavior, in 2D as well as
in 3D models. The result which is the most relevant for us is the difference
of 30 per cent on the predicted mean equivalent width $\langle W \rangle$ 
of the line, leading to a similar change in the derived lithium abundance
(NLTE/LTE abundance correction +0.15 dex). But another interesting difference
is the  reverse behavior of the equivalent width $W$ as a function of surface
continuum  brightness $I_c$. While LTE computations result in a strongly
positive slope  in the $W$ versus $I_c$ diagram (with a large scatter around
the mean  relation), NLTE computations show a slightly negative slope and a
much tighter  (anti-)correlation between $I_c$ and $W$. This reflects the fact
that the  population of the Li~I levels is much more controlled by the local
temperature  in LTE than in NLTE, where, for weak lines, the photoionization
rates play the  dominant role.
So, even if the general conclusion of the above mentioned papers is that, 
in the solar case, the abundance determination of Li is not strongly affected 
by the combined effects of temperature fluctuations and NLTE, in comparison to 
what is obtained with classical 1D models having the same effective temperature
and gravity, it appears unsafe to compute abundances from multidimensional 
stellar atmospheres based on the assumption of LTE line formation.

\section{2D radiation hydrodynamics models}

\begin{figure}
\vspace*{-100mm}
\plotfiddle{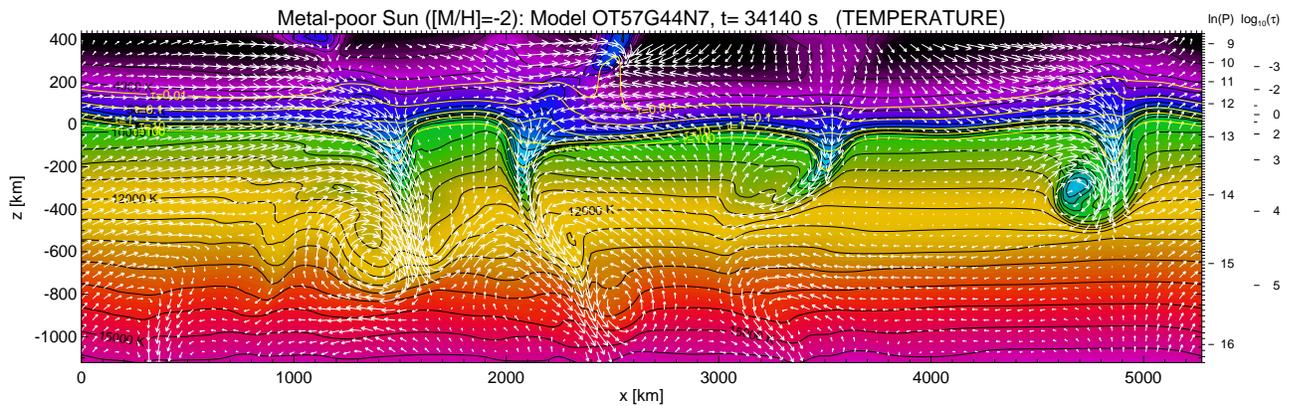}{15cm}{90}{65}{65}{300}{-30}
\caption{Snapshot from the metal-poor Sun simulation, showing the velocity 
field (arrows) and temperature structure (black contour lines).
White lines are curves of iso-optical depth. The origin of the geometrical
scale (left) corresponds to $\tau_{\rm Ross}=1$; scales at right refer to
average gas pressure $P$ and average optical depth $\tau_{\rm Ross}$. 
This simulation was done on a 210x106 grid (tick marks at right and top),
with a vertical extent of about 1\,500~km ($\approx 7 H_p$), and a horizontal 
period of 5\,250~km (upper and lower boundaries `open', lateral boundaries 
periodic).}
\end{figure}  

Our LTE and NLTE computations have been performed on the basis of several 
snapshots from a 2D numerical simulation of convection in a stellar envelope 
having the same effective temperature and gravity as the Sun, but a 100 times 
smaller metallicity. Basically, the time dependent equations of hydrodynamics 
are solved for a compressible fluid, with an energy equation including 3 terms:
turbulent and shock dissipation of kinetic energy, diffusive transport of
heat, and radiative energy exchange. The main limitation of the 
code is the restriction of the flow to two spatial dimensions.
Magnetic fields and rotation are ignored.

Apart from these simplifications, as much realistic physics as
possible is included. The equation of state accounts for ionization of 
hydrogen and helium as well as H$_2$ molecule formation, opacities 
have been adapted from Kurucz's ATLAS code and include line absorption.
For the computation of the radiative energy balance, we employ a
multi-dimensional, non-local, frequency-dependent radiative transfer scheme, 
actually solving the transfer equation along 26880 independent rays of 
various inclinations, using an efficient modified Feautrier method 
(Feautrier 1964). At the bottom boundary, inflowing matter has a given 
specific entropy, which is adjusted to produce the prescribed effective 
temperature of the atmosphere. Energy dissipation on small scales is roughly 
modeled by introducing a subgrid scale eddy viscosity, depending on the grid 
resolution and local velocity gradients in the usual way.  
Details of the employed hydrodynamics code can be found 
in Ludwig et al.\ (1994) and Freytag et al.\ (1996).

Fig.~1 shows a sample snapshot from our metal-poor Sun simulation. Note the 
complex velocity pattern and the occurrence of very strong temperature 
gradients. The relevant region for the formation of the Li~I line is the
$\tau = 0.1 $ contour line.

\section{ NLTE computation of the Li~I resonance line}

The computation of the Li~I spectrum is greatly simplified by the fact that all 
lines of Li~I are weak, Li being a trace element. So the radiation field in the 
line is, to first approximation, the same as the continuous radiation
field, a single iteration being sufficient for taking care of the small 
perturbation of the monochromatic radiation field brought about by the line.
        
We have, as a first step, approximated the Li~I atomic configuration by a five 
level atom, exactly as done by Uitenbroek (1998). This leaves six 
permitted bound-bound transitions, and five photoionization rates needing
the computation of the continuous radiation field at frequencies above the 
threshold, until the contribution of the product of photoionization 
cross-section and mean intensity of the radiation becomes negligible. 
Because in the UV the contribution of lines to the opacity is important, 
we have used Kurucz's Opacity Distribution Functions (ATLAS~9)
for the relevant metallicity. This multiplies the computation time by 12, 
as each opacity bin is subdivided into 12 subintervals. So, for each snapshot, 
the transfer equations must be solved for about $12 \times 120$ wavelengths, 
along 26880 different rays (note that these extensive computations are done 
only after the actual hydrodynamical simulation for a few selected snapshots).
After this, it is possible to compute all the coefficients in the equations of 
statistical equilibrium (see Mihalas 1970, p. 144). Once the departure 
coefficients $b_i$ are evaluated, the line can be computed using
the source function:
$$S=\frac {\kappa_c}{\kappa_l+\kappa_c} B_{\nu} + \frac{\kappa_l} 
    {\kappa_l+\kappa_c}S_l $$
where:
$$S_l = S_{ij} = \frac{2h\nu_{ij} ^3}{c^2}  
        \frac{1}{(b_i/b_j)\exp(h\nu_{ij} /kT)-1} $$
is the line source function. The departure coefficient for the lower level is 
$b_i$ and the one for the upper level $b_j$. The other notations are standard. 
Subscript ``c'' stands for continuum, and ``l'' for ``line''.
Note that in NLTE the expression of the partition function is modified and
becomes:

$$U=\sum_{i}b_i g_i \exp(-\chi_i /kT) $$
where the $g_i$ and $\chi_i$ are the statistical weight and the excitation 
energy of level $i$, respectively .

\section{Results and discussion}

We have first tested our program on a Kurucz ATLAS~9 1D solar model to see 
whether the $b_i$ had the expected behavior, already computed by Carlsson 
et al.\ (1994). Fig.~2 shows the depth-dependence of the first 3 departure
coefficients, applying to levels 2s, 2p and 3s, respectively.

\begin{figure}[h]
\vspace*{-10mm}
\plotfiddle{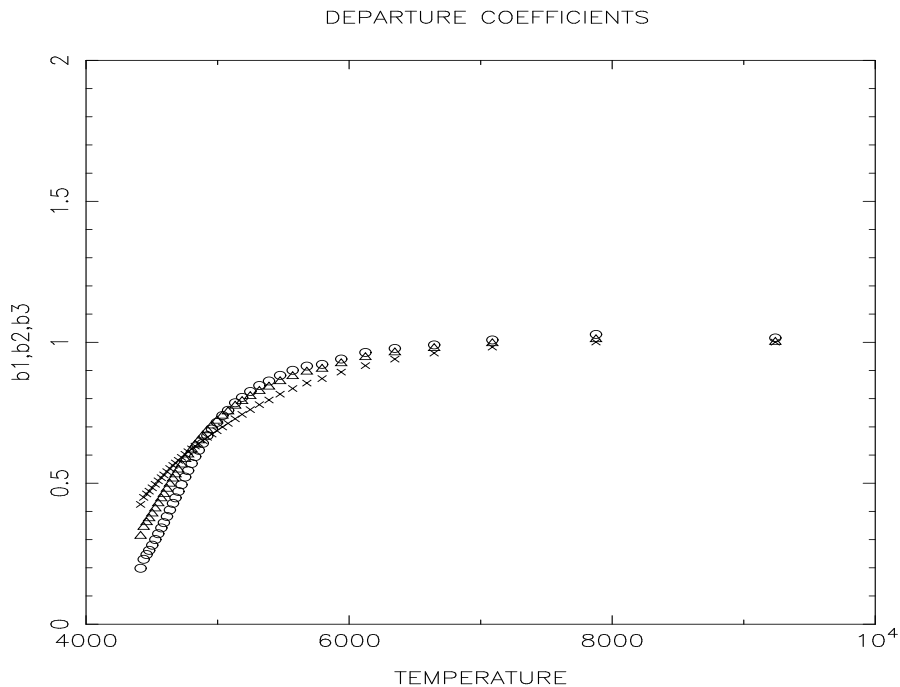}{10cm}{0}{70}{50}{-220}{-85}
\caption{ Variation of the first three departure coefficients with depth 
(actually temperature in this graph), based on an 1D ATLAS~9 model of the
solar atmosphere (mixing-length parameter $\alpha=0.5$). Circles refer to 
the ground level ($b_1$), triangles to the 2p level ($b_2$), and crosses 
to the 3s level ($b_3$). As expected, all three coefficients become very 
close to 1 in the deep layers, and reach values below 0.5 in surface layers, 
as found by Carlsson et al.\ (1994). The adopted logarithmic abundance of Li 
is 2.2, on the scale log(nH)= 12.0.}  
\end{figure}

\begin{figure}[!t]
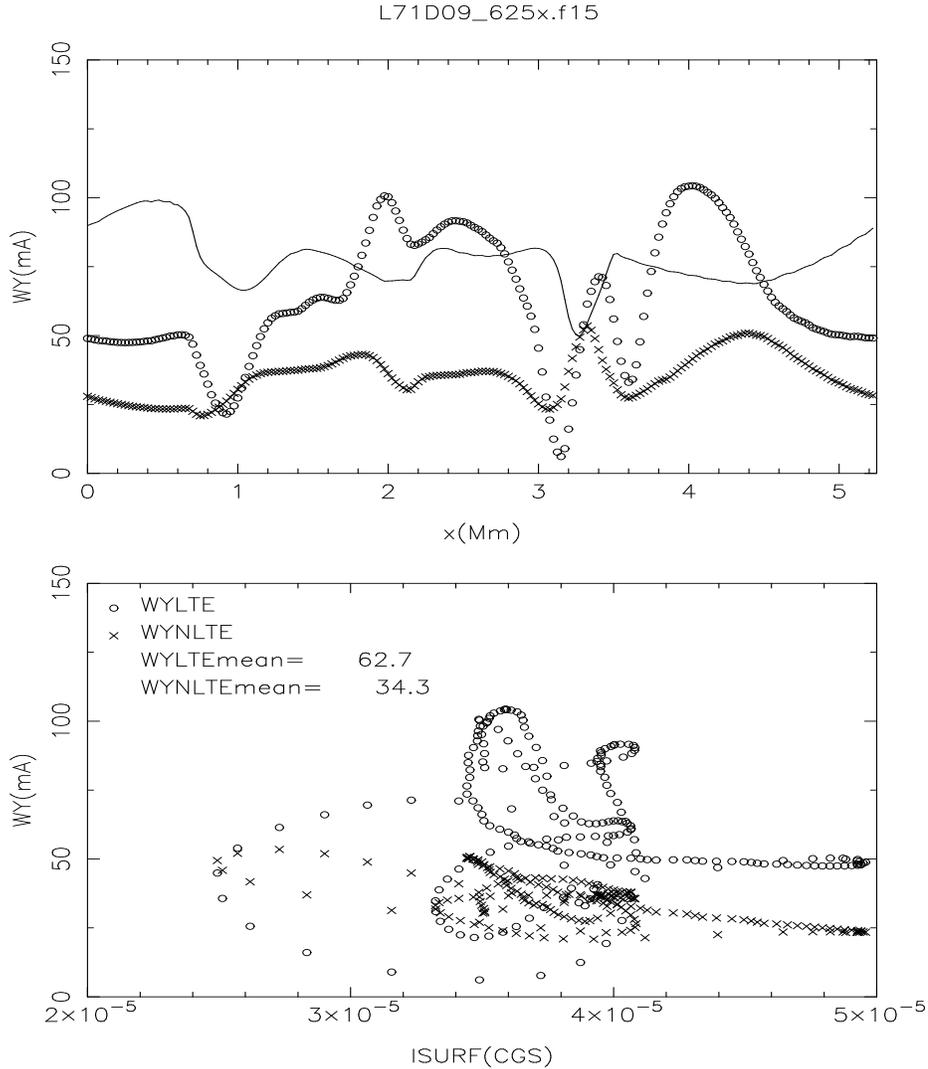

\vspace*{-5mm}
\plotfiddle{cayFig3a.eps}{70mm}{0}{70}{55}{-230}{-100}
\plotfiddle{cayFig3b.eps}{65mm}{0}{70}{55}{-230}{-100}
\caption{{\bf Top:} LTE (circles) and NLTE (crosses) equivalent widths of the 
Li~I resonance line as a function of horizontal position $x$, for a
representative snapshot from a 2D simulation of {\em solar granulation\/}. 
The thin line gives the continuum intensity at $\lambda$~670.8 nm in arbitrary 
units. {\bf Bottom:} Same equivalent widths as a function of continuum 
intensity ISURF. All data derived from vertical rays ($\mu$=1).} 
\end{figure}

Next, we have computed the equivalent width of the Li~I resonance line for 
the Kurucz 1D solar model and for two 2D solar snapshots, still for the 
logarithmic Li abundance 2.2. Fig.~3 shows the variation of the equivalent 
width $W$ (for the {\em intensity\/} normal to the surface) of the 
Li~I 670.8~nm line over the simulated granulation pattern, both as a function 
of the horizontal position $x$ (top) and as a function of the continuum 
intensity $I_c$ (bottom) for one particular snapshot. 
Note the wide variation of $W$ computed in LTE, compared to the much
more limited excursion of $W$ computed in NLTE. The mean equivalent widths 
for the {\em flux\/} spectrum integrated over the full length of the sample 
are given in Table~\ref{T1} for the two solar snapshots and for the 1D 
reference model having the same effective temperature, gravity and (solar) 
metallicity. In each case, the line is computed in LTE as well as in NLTE. 
We note that, in NLTE, the results for the 2D snapshots do not differ 
significantly from the 1D case.

\begin{table}[t]
\caption{LTE and NLTE mean equivalent width $\langle W \rangle$ [m\AA] of the 
Li~I $\lambda$~670.8 line for 2 different snapshots from a 2D hydrodynamical 
simulation of {\em solar surface convection\/}, obtained from the horizontally 
averaged {\em flux\/} spectrum. For comparison, the equivalent widths resulting 
for a 1D ATLAS9 reference model, computed with the same line formation code, 
are given in the first row. The results from Kieslman (1997) given in the last
rows have been rescaled to the same Li abundance of 2.2; they refer to 
intensity ($\mu=1$).}
\begin{center}
\begin{tabular}{ccc}
\tableline
1D model or RHD snapshot & $\langle W \rangle$ (LTE) & $\langle W \rangle$ 
(NLTE) \\
\tableline
1D Kurucz ($\alpha=0.5$) & 43 & 32 \\
2D, L71D09-605 & 59 & 35 \\
2D, L71D09-625  &  71  & 36 \\
\tableline
Kiselman 1D (OSMARCS) & 44  & 42 \\
Kiselman 3D           & 52  & 37 \\
\tableline\tableline
\end{tabular}
\end{center}
\vspace*{-5mm}
\label{T1}
\end{table}

\begin{figure}[!t]
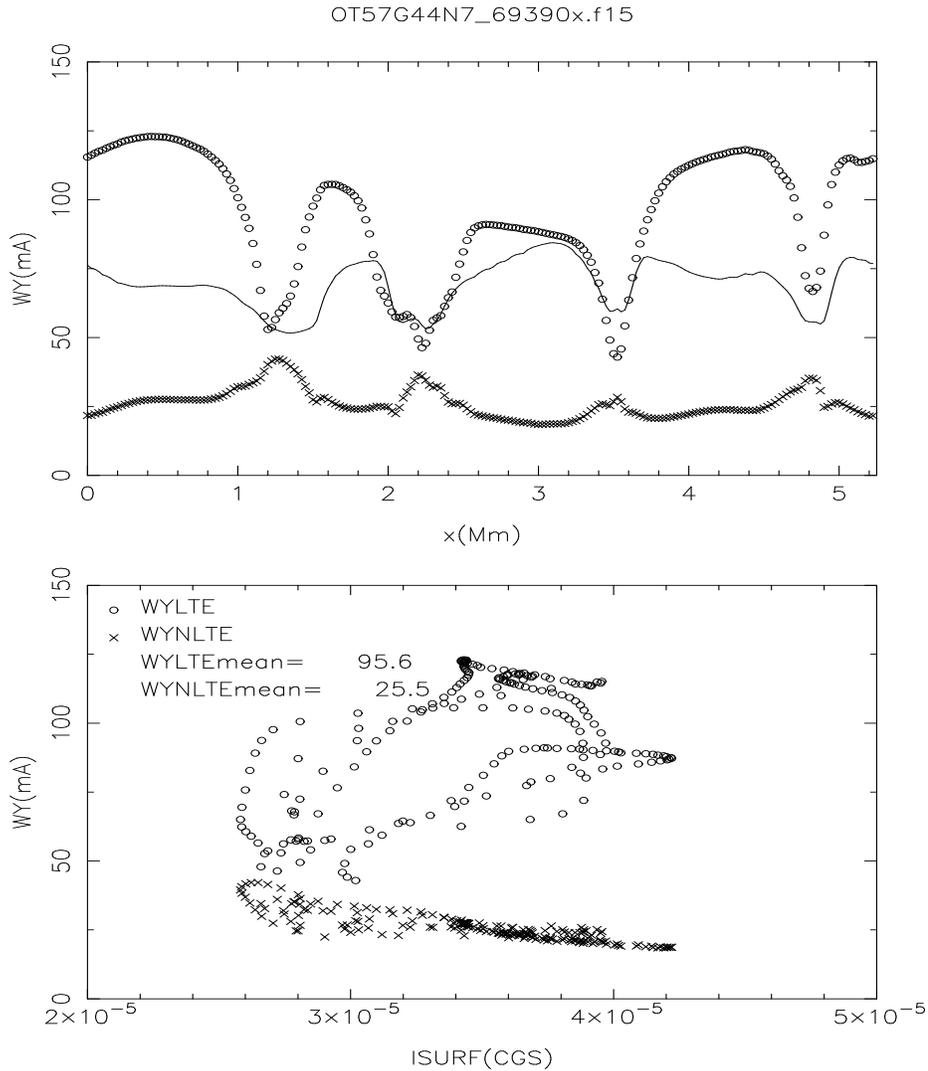

\vspace*{-5mm}
\plotfiddle{cayFig4a.eps}{70mm}{0}{70}{55}{-230}{-100}
\plotfiddle{cayFig4b.eps}{65mm}{0}{70}{55}{-230}{-100}
\caption{ Same plots as Fig.~3, but for the snapshot from the
{\em metal-poor star simulation\/} shown in Fig.~1} 
\end{figure}

Finally, we have carried out a similar procedure for our metal-poor stellar
example, again computing the equivalent width of the Li~I resonance line for 
a Kurucz 1D model and for five snapshots from our 2D hydrodynamical
simulation; as before, a Li abundance of 2.2 was adopted. Fig.~4 shows the 
variation of the equivalent width $W$ of the Li~I resonance line over the 
stellar granulation pattern for the typical snapshot displayed in Fig.~1. 
The difference between LTE and NLTE is even more pronounced than
in the solar case, the NLTE correlation between $W$ and $I_c$ being much
tighter and of opposite sign compared to LTE. 

\begin{table}[!h]
\caption{ Same as Table~\ref{T1}, but for 5 different
snapshots from a 2D hydrodynamical simulation of surface convection in
a {\em metal-poor star\/} ($[M/H]=-2$) with solar effective temperature and 
gravity. For comparison, the results for the corresponding 1D ATLAS9 model, 
computed with the same line formation code, are given in the first row. 
}
\begin{center}
\begin{tabular}{ccc}
\tableline
1D model or RHD snapshot & $\langle W \rangle$ (LTE) & $\langle W \rangle$ 
(NLTE) \\
\tableline
1D Kurucz ($\alpha=0.5$) &  ~38.4 & 26.1 \\
T57G44N7-43108 & 116.0 & 27.7 \\
T57G44N7-69390 & 115.4 & 27.1 \\
T57G44N7-83752 & 112.7 & 27.3 \\
T57G44N7-87007 & 110.9 & 27.7 \\
T57G44N7-97644 & 108.2 & 26.5 \\
\tableline\tableline 
\end{tabular}
\end{center}
\vspace*{-5mm}
\label{T2}
\end{table}

The mean equivalent widths derived from the horizontally averaged
{\em flux\/} spectrum are listed in Table~\ref{T2} for the 5 snapshots 
and for the Kurucz 1D reference model. In LTE, the 1D/2D difference is huge
(granulation abundance correction $\approx -0.45$~dex). But remarkably,
the 2D NLTE line strengths show very little dispersion and
do not indicate any significant offset with respect to the 1D case. 
An obvious conclusion from these results is that the 2D LTE computations 
are way off, strongly underestimating the Li abundance.
In NLTE, the error introduced by representing the inhomogeneous stellar
atmosphere by a flux-constant 1D Kurucz model appears to be almost negligible. 

The mechanism behind the spatial variation of the line strength is clearly 
identified on Figs~1 and 4: 
hot granules produce at the same time a steeper temperature gradient 
and a lower temperature in the line formation region. In LTE, the latter 
leads to an overpopulation of the lower level of the transition due to a 
shift of the ionization equilibrium towards neutral particles (Saha equation). 
Since both effects enhance the equivalent width of the line, the correlation 
between continuum intensity and LTE line strength is clearly positive 
(there is considerable dispersion around the mean $W(I_c)$ relation, however,
due to the presence of inclined thermal inhomogeneities). This result is 
inverse of what is expected for a set of hot and cool {\em radiative 
equilibrium atmospheres} lined up side by side: here the line would weaken in 
the hot atmosphere, because the population of the lower level is the dominant 
factor. It is clear that the reasoning of Kurucz fails, essentially because 
the actual vertical stratification of hot and cold regions has little to do 
with the stratification of hot and cool radiative equilibrium atmospheres.

A positive $W$-$I_c$ correlation alone is not sufficient to explain the LTE 
result that the mean equivalent width is much larger in 2D than in 1D:
as long as the fluctuations of the line strength with temperature remain in 
the linear domain, the mean equivalent width is not affected by the temperature
fluctuations. But as the population of the ground level $N_0$ of Li I varies 
{\em exponentially} with T, a nonlinearity sets in: symmetrical 
temperature fluctuations produce asymmetrical population variations.
Assume that Li is almost completely ionized and consider only the exponential 
factor of the Saha equation for simplicity. 
Then 
$$N_0(T)=N_0(T_0)\,\exp(+\chi/k T)=N_0(T_0)*\left\{1 - q s + q s^2 + \frac{1}{2}
  q^2 s^2 + \ldots \right\} $$
where $\chi$ is the ionization potential and we have defined 
$q \equiv \chi/k T_0$ and $s \equiv (T-T_0)/T_0$, $T_0$ being the mean 
temperature. Taking the horizontal average assuming  symmetrical temperature
variations ($\langle s \rangle = 0$), we obtain $\langle N \rangle \approx 
N_0(T_0)*(1 + 0.5\, q^2 \langle s^2 \rangle$ ($\sqrt{\langle s^2 \rangle} 
\le 0.04$; $q \approx 12$).
So the mean of $W$ is biased towards larger equivalent widths in an 
inhomogeneous atmosphere. This explains part of the 1D/2D difference
found in our numerical computations. The main contribution, however, 
is attributed to the lower mean photospheric temperature in the 2D
model as a result of adiabatic cooling due to overshooting: $T_0(2D) < T(1D)$.

In NLTE, the local temperature plays little role. Rather, the radiation field 
is the dominant factor. Hence, photoionization overionizes the lower level 
over a hot granule, and the equivalent width becomes smaller.  
As a result, NLTE equivalent widths are anti-correlated with the continuum
surface brightness. The variation of $W$ is much smaller than in LTE, because
the angle-averaged radiation field depends only weakly on horizontal position
at the height of line formation. 

Finally, we would like to mention the work by M. Asplund, who has also
presented his NLTE line formation calculations for Li~I on this symposium.
His results are based on the 3D snapshots that he had used previously for 
the LTE investigation of the subgiants HD 140283 and HD 84937.
He reached, on this completely independent set of hydrodynamical models, and 
with a different NLTE code, the same conclusions as we did: NLTE line formation
in multidimensional models is quite different from the LTE case, in the way 
that the NLTE multi-dimensional Li abundance is much closer to the abundance
derived from classical flux-constant 1D models. A detailed comparison of our 
results shows that the remaining differences can be traced to the use of a 
Li~I atom with 5 levels in our case, and with 20 levels in the case of Asplund 
et al.\ (this volume). Another difference is that the temperature inhomogeneities
are somewhat enhanced in 2D models with respect of what occurs in 3D models,
leading to correspondingly larger LTE granulation abundance corrections.
But this is only a minor point. Certainly, the two groups agree that LTE 
Li abundance determinations relying on multidimensional hydrodynamical
simulations of convection in metal-poor dwarfs are highly misleading.

\section{Conclusions}

1. The statement of Kurucz (1995) that abundances of lithium derived from 
standard 1D models of metal-poor stellar atmospheres is too small by a 
factor of 10 is not supported by actual multidimensional NLTE computations. 
Even the sign of the correction is doubtful, and the error is well below 
0.1 dex, both according to our investigation and the one presented by 
M. Asplund on this symposium. \\

\noindent
2. LTE abundance determinations based on inhomogeneous atmospheres are strongly 
discouraged. They produce large ``granulation abundance corrections'' 
due to non-linear effects in the direction opposite to Kurucz' prediction, 
but the actual NLTE line formation mechanism couples the population
of the atomic levels more closely to the mean radiation field than to
the local temperature. \\

\noindent
3. The combination of multidimensional models with NLTE line formation  
for the Li~I $\lambda$~670.8 nm resonance line leads to the same lithium
abundance as that derived from NLTE analysis with flux-constant 1D models,
abundance differences being less than 0.1 dex. However, this result must 
not be hastily generalized to other atoms with a different atomic 
structure. \\

\noindent
4. An obvious future improvement is to extend the NLTE analysis to the 
continuum, which has be assumed here to be in LTE. We plan to do that in 
the near future. If it turns out that the H$^-$ ion is affected by NLTE, 
this will raise  a new question: should such effects be included 
already in the radiation hydrodynamics code, which determines
the amplitude of the thermodynamical fluctuations?

\end{document}